\documentclass[prl,twocolumn,superscriptaddress]{revtex4}
\usepackage{amssymb,amsmath,amsthm,graphicx}
\usepackage{latexsym}
\usepackage{bm}
\usepackage[]{hyperref}
\usepackage{cleveref}

\pdfoutput=1

\def\ie{{\it i.e. }}

\begin{document}

\title{Lattice Black Branes: Sphere Packing in General Relativity}

\author{\'Oscar~J.~C.~Dias}
\email{ojcd1r13@soton.ac.uk}
\affiliation{STAG research centre and Mathematical Sciences, University of Southampton, UK} 
\author{Jorge~E.~Santos}
\email{jss55@cam.ac.uk}
\affiliation{Department of Applied Mathematics and Theoretical Physics, University of Cambridge, Wilberforce Road, Cambridge CB3 0WA, UK} 
\author{Benson~Way}
\email{benson@phas.ubc.ca}
\affiliation{Department of Physics and Astronomy, University of British Columbia, 6224 Agricultural Road, Vancouver, B.C., V6T 1W9, Canada} 

\begin{abstract}
We perturbatively construct asymptotically $\mathbb{R}^{1,3}\times\mathbb{T}^2$ black branes with multiple inhomogeneous directions and show that some of them are thermodynamically preferred over uniform branes in both the microcanonical and canonical ensembles.  This demonstrates that, unlike five-dimensional black strings, the instability of some unstable black branes has a plausible endpoint that does not require a violation of cosmic censorship.
\end{abstract}

\maketitle

{\bf~Introduction --} Higher-dimensional black holes are markedly different from those in four-dimensions. In particular, they can be unstable \cite{Gregory:1993vy,Dias:2009iu,Dias:2010eu,Hartnett:2013fba,Dias:2014eua,Santos:2015iua}, violate black hole uniqueness \cite{Emparan:2001wn,Dias:2009iu,Dias:2010eu,Dias:2014cia} and also lead to a violation of weak cosmic censorship \cite{Lehner:2010pn,Figueras:2015hkb,Figueras:2017zwa}. 

The prototypical representative of such behaviour is the uniform five-dimensional black string, which is the product space of a Schwarzschild black hole with a circle.  The black hole therefore has spatial horizon topology $S^2\times S^1$, and is asymptotically $\mathbb R^{1,3}\times S^1$. When the mass of the black string is much smaller than the size of the circle, the $S^2$ extent of the horizon is much smaller than the $S^1$ length. Such a separation of length scales leads to an instability that was first discovered by Gregory and Laflamme in \cite{Gregory:1993vy}.

What is the endpoint of this instability?  When first proposed by \cite{Gregory:1993vy}, there were two candidates: a nonuniform black string or a localised black hole.   Nonuniform blacks strings have the same horizon topology as uniform black strings, but they no longer respect the symmetries of the $S^1$. The nonuniform solutions branch off from uniform solutions at a zero mode located at the critical onset of the Gregory-Laflamme (GL) instability. A localised black hole has spherical horizon topology $S^3$. They were expected to exist since it is possible to place a small spherical black hole within $\mathbb R^{1,3}\times S^1$. The possibility of evolution towards localised black holes was especially intriguing because the change in horizon topology implies a violation of weak cosmic censorship.

Later results \cite{Gubser:2001ac,Wiseman:2002zc} ruled out the nonuniform strings as an endpoint since they have lower entropy (horizon area) than the uniform strings, while  localised black holes remained a possibility since they are entropically preferred \cite{Kudoh:2004hs}.  The time evolution of the instability was finally performed in \cite{Lehner:2010pn}, providing the first numerical evidence of a violation of weak cosmic censorship.

The Gregory-Laflamme instability is a generic feature of black objects with extended directions.  The instability occurs in asymptotically flat black holes, as well as theories with compact directions such as those in string theory and AdS/CFT \cite{Maldacena:1997re,Gubser:1998bc,Witten:1998qj,Aharony:1999ti}. However, unlike the black string, most of these scenarios contain multiple extended directions. The situation is therefore more complex since there is a larger space of unstable perturbations, and consequently a larger space of nonuniform solutions. We wish to shed light on which of these perturbations lead to the most entropically favourable nonuniform solutions, and whether these solutions can be dominant over uniform solutions. 

Consider then the six-dimensional black brane
\begin{equation}\label{uniformbrane}
\mathrm ds^2=-\left(1-\frac{r_0}{r}\right)\mathrm dt^2+\frac{\mathrm dr^2}{1-\frac{r_0}{r}}+r^2\mathrm d\Omega^2_2+\mathrm dx_1^2+\mathrm dx_2^2\;,
\end{equation}
where $r_0$ is the horizon radius. The linear stability calculation for this system is identical to the black string \cite{Gregory:1993vy}.  Taking a metric perturbation \emph{ansatz} $g_{\mu\nu}=g^{(0)}_{\mu\nu}+e^{-i\omega t+ik_1 x_1+ik_2 x_2}h_{\mu\nu}$, one finds that for fixed $r_0$, the frequency $\omega$ depends only on the norm $k_1^2+k_2^2$.  In particular, the zero mode $\omega=0$ occurs at a critical $k_{GL}=\sqrt{k_1^2+k_2^2}$.  Each frequency, including the zero mode, therefore contains a circle of degenerate perturbations.  Because of this degeneracy, it is possible for multiple static solutions to appear from the same zero mode. 

To restrict some of these possibilities, let us take the $x_1$ and $x_2$ directions to be identified on an oblique torus with a single length scale $L$
\begin{equation}\label{x1x2iden}
x_1\sim x_1+(n_1+\gamma n_2)L\;,\qquad x_2\sim x_2+n_2\sqrt{1-\gamma^2} L\;,
\end{equation}
for any integers $n_1$ and $n_2$ and angle $\gamma\in[0,1)$.

Our goal is to perform a perturbative analysis of nonuniform black branes with these toroidal symmetries and compute their thermodynamic quantities.  We will find that it is possible for these solutions to be thermodynamically favoured over uniform solutions.  

As an aside, we point out that there is another situation where nonuniform solutions can be entropically favoured. For $D$-dimensional black strings consisting of a $D-1$ dimensional Schwarzschild-Tangherlini black hole and a circle, nonuniform black strings are favored for $D\geq 14$ \cite{Sorkin:2004qq,Figueras:2012xj}.  Time evolution within a large $D$ expansion indicates that evolution proceeds towards the nonuniform solution, avoiding a violation of cosmic censorship.  Black branes on oblique torii were also studied at lowest order in large $D$ \cite{Rozali:2016yhw}, but unfortunately different static solutions have the same thermodynamic quantities at this order. 

We also mention that related holographic lattice solutions in $AdS_4$ were studied in \cite{Donos:2015eew}, where a triangular lattice is thermodynamically preferred. 

\begin{figure}
\centering
\includegraphics[width=.4\textwidth]{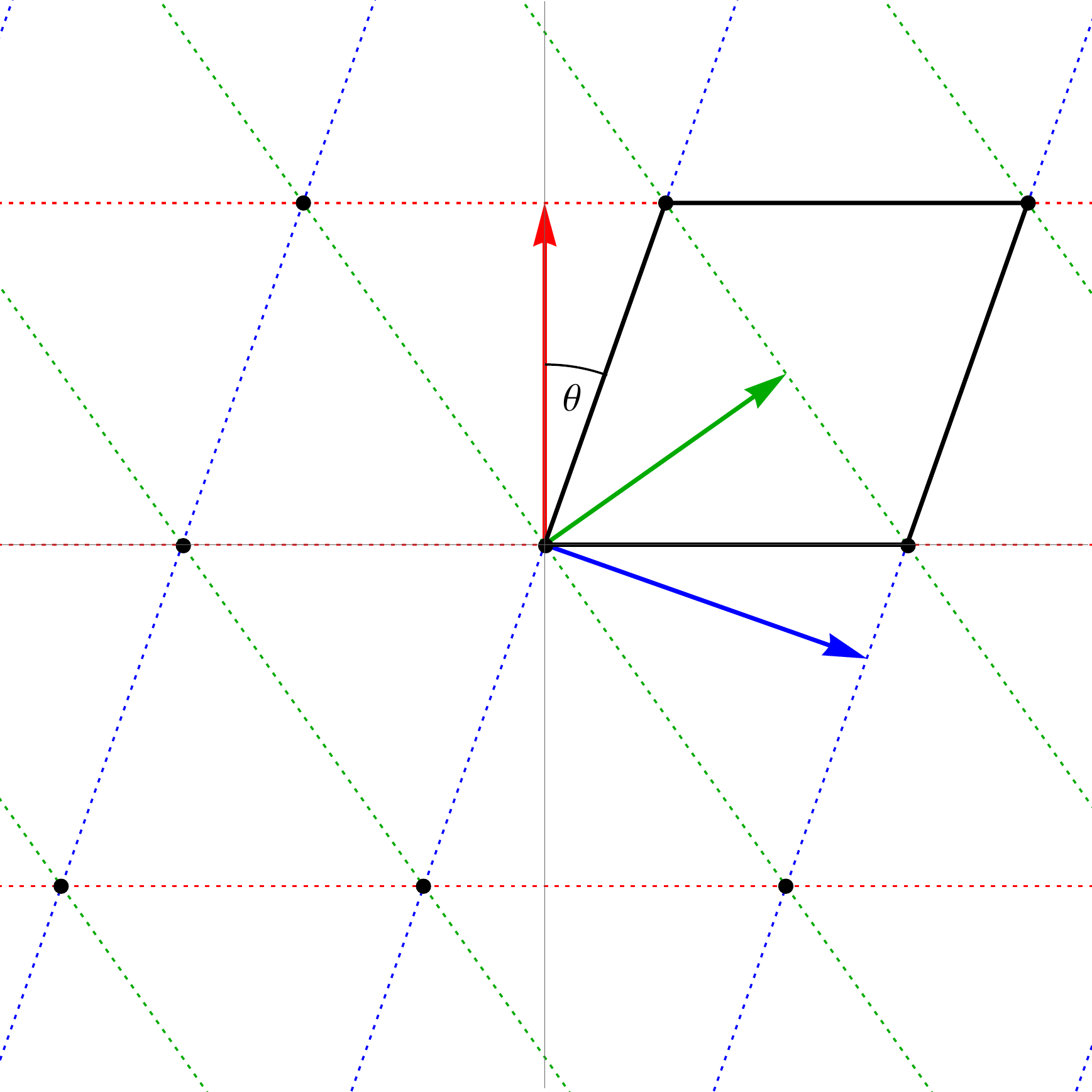}
\caption{The fundamental torus region is given by the solid black parallelogram with side lengths $L$, which is angled at $\theta=\arcsin\gamma$.  The arrows show the longest wavelength wavevectors $2\pi\vec k/|k|^2$ that generate parallel lines in the lattice.}\label{fig:torus}
\end{figure}

{\bf~Perturbative and numerical methods --}
The metric perturbations only depend on the norm $k_1^2+k_2^2$, but the torus boundary conditions do not allow arbitrary wavevectors. Each wavevector $\vec k=(k_1,k_2)$ describes a plane wave which must fit within the torus. The allowable wavevectors with the longest wavelengths can be viewed in Fig. \ref{fig:torus} where the torus has been extended to a periodic lattice spanning $\mathbb R^2$. The parallel lines on this lattice allow us to visualise the permissible plane waves with the longest wavelengths.  We see that the torus generates three sets of parallel lines, corresponding to three different wavevectors. By demanding that the torus has a single length scale $L$, two of these wavevectors must have the same norm. The third wavector also has this norm only when $\gamma=1/2$. 

Let us see this more explicitly in terms of the perturbations.  We first move to a set of coordinates that are adapted to the torus:
\begin{equation}\label{12toxy}
x_1=\frac{L}{2\pi}(x+\gamma y)\;,\qquad x_2=\frac{L}{2\pi}\sqrt{1-\gamma^2}y\;,
\end{equation}
where the torus identifications \eqref{x1x2iden} imply that
\begin{equation}\label{xyiden}
x\sim x+2\pi n_x\;,\qquad y\sim y+2\pi n_y\;.
\end{equation}
In these coordinates, a static metric perturbation must be of the form $g_{\mu\nu}=g^{(0)}_{\mu\nu}+e^{i n_x x+i n_y y}h_{\mu\nu}$, for integers $n_x$ and $n_y$.  Mapping that back into $x_1$ and $x_2$ coordinates, we can identify the wavevector components $k_1$ and $k_2$.  The norm of the wavevector is given by
\begin{equation}\label{keq}
|k|=\sqrt{k_1^2+k_2^2}=\frac{1}{\lambda}\sqrt{n_x^2+n_y^2-2n_x n_y \gamma}\;,
\end{equation}
where we have defined the length scale 
\begin{equation}\label{lambdadef}
\lambda=\frac{\sqrt{1-\gamma^2}}{2\pi}L.
\end{equation} 
When $L$ is sufficiently small relative to $r_0$, there is no instability. Now consider increasing $L$.  Since the uniform black brane is unstable for perturbations with $|k|<k_{\mathrm{GL}}$, the smallest $|k|$ (longest wavelength) perturbations will be the first to generate static nonuniform branes.  We therefore wish to determine which integers $n_x$ and $n_y$ generate the smallest $|k|$ perturbations.  These are
\begin{equation}\label{pert1}
n_x=\pm 1\;, n_y=0;\qquad\text{and}\qquad n_x=0\;, n_y=\pm 1\;,
\end{equation}
which have $|k|=1/\lambda$, and
\begin{equation}\label{pert2}
n_x=n_y=\pm1\;,
\end{equation}
which has $|k|=2(1-\gamma)/\lambda$. Both sets of perturbations have the same $|k|$ precisely at $\gamma=1/2$.  Naturally, there are shorter wavelength perturbations than those in \eqref{pert1} and \eqref{pert2}.  These perturbations only become unstable for larger torii, and we do not consider them here.

At this point, we would like to distinguish between two types of perturbations that generate static nonuniform solutions: those that have a single wavevector, and those that are a linear combination of different wavevectors.  Let us call nonuniform branes generated by the former as ``semi-nonuniform" black branes, and those by the later as ``fully nonuniform" black branes.  Semi-nonuniform black branes preserve translation invariance in one direction and are therefore equivalent to the product of a 5-dimensional nonuniform black string and an extra flat direction.  Fully nonuniform black branes have no such symmetry. 

In this work, we focus on the fully nonuniform black branes. For $\gamma\neq1/2$, such black branes can be generated by the two perturbations in \eqref{pert1}. For $\gamma=1/2$, it turns out that fully nonuniform branes can only come from a perturbation with all three wavevectors in \eqref{pert1} and \eqref{pert2}.  These fully nonuniform black branes will compete with both the uniform black brane and semi-nonuniform black branes. However, we know from previous results \cite{Gubser:2001ac,Wiseman:2002zc} that semi-nonuniform black branes in this dimension are never thermodynamically preferred over the uniform phases. 

Now let us proceed with the gravitational calculation. We will work in the $x$ and $y$ coordinates as well as a new radial coordinate from the transformation
\begin{equation}
r=\frac{r_0}{1-z^2}\;.
\end{equation}
Our metric \emph{ansatz} is given by
\begin{align}\label{ansatz}
{\mathrm ds}^2=r_0^2&\bigg\{ -z^2 q_1 \frac{{\mathrm d}t^2}{r_0^2}+\frac{4  q_2 {\mathrm d}z^2}{(1-z^2)^4}\, +\frac{ q_3}{(1-z^2)^2}\,{\mathrm d}\Omega^2_{S^2}\nonumber\\
&\quad + \overline\lambda^2\bigg[\frac{q_4}{1-\gamma^2} \left({\mathrm d}x + \gamma q_6 {\mathrm d}y +\frac{q_7}{\overline\lambda} {\mathrm d}z \right)^2\nonumber\\
&\qquad\qquad\qquad+q_5\left({\mathrm d}y+\frac{q_8}{\overline\lambda} {\mathrm d}z \right)^2
\bigg]\bigg\},
\end{align}
where $q$'s are functions of $\{x,y,z\}$, and we have also defined $\overline\lambda=\lambda/r_0$.  Note that implicit in this \emph{ansatz} is a horizon at $z=0$ and asymptotic infinity at $z=1$.  The uniform black brane \eqref{uniformbrane} can be obtained in these new coordinates by setting $q_i(z,x,y)=\bar{q}_i$ with
\begin{equation}\label{unif}
\bar{q}_{1,2,3,4,5,6}=1\;,\qquad \bar{q}_{7,8}=0\;.
\end{equation}

The eight functions $q_i$ are obtained by solving the vacuum Einstein equation $R_{\mu\nu}=0$, subject to the appropriate boundary conditions. We will do this perturbatively about the black brane solution stopping at the order where thermodynamic quantities become corrected. 

We use the Einstein-DeTurck formalism \cite{Headrick:2009pv,Figueras:2011va,Wiseman:2011by,Dias:2015nua}, which is valid non-perturbatively as well. This method requires a choice of reference metric $\overline g$, which contains the same symmetries and causal structure as the desired solution.  The reference metric we choose is the black brane metric given by \eqref{unif}. The DeTurck method modifies the Einstein equation $R_{\mu\nu}=0$ into
\begin{equation}\label{EdeT}
R_{\mu\nu}-\nabla_{(\mu}\xi_{\nu)}=0\;,\qquad \xi^\mu \equiv g^{\alpha\beta}[\Gamma^\mu_{\alpha\beta}-\overline{\Gamma}^\mu_{\alpha\beta}]\;,
\end{equation}
where $\Gamma$ and $\overline{\Gamma}$ define the Levi-Civita connections for $g$ and $\bar g$, respectively. Unlike $R_{\mu\nu}=0$, this equation yields a well-posed elliptic boundary value problem. It was proved in \cite{Figueras:2011va} that static solutions to \eqref{EdeT} necessarily satisfy $\xi^\mu=0$, and hence are also solutions to $R_{\mu\nu}=0$.

We now discuss the boundary conditions. At $z=1$, the solution must be asymptotically $\mathbb{R}^{1,3}\times\mathbb{T}^2$ which requires $q_i{\bigl |}_{z=1}=\bar{q}_i$.  At the horizon $z=0$, we require regularity and impose $q_1{\bigl |}_{z=0}=q_2{\bigl |}_{z=0}$, $\partial_z q_i{\bigl |}_{z=0}=0$ for $i=2,\ldots, 6$, and $q_7{\bigl |}_{z=0}=0=q_8{\bigl |}_{z=0}$. Finally, we impose periodic boundary conditions on $x$ and $y$ according to \eqref{xyiden}.

Since $r_0$ is just an overall scale, our \emph{ansatz} \eqref{ansatz} is parametrised by $\overline\lambda$ and $\gamma$. However, it's more convenient in our perturbative calculation to replace $\overline\lambda$ by $|k|$.  $\overline\lambda$, $\lambda$ or $L$ can be recovered via \eqref{keq}, \eqref{lambdadef}, and $\overline\lambda=\lambda/r_0$.

Now expand the metric functions and $|k|$ in powers of $\epsilon$:
\begin{eqnarray}\label{expansion}
\hspace{-0.3cm} q_i=\bar{q}_{i}+\sum_{n=1}^{\infty} \epsilon^n \,q_i^{(n)}(z,x,y), \quad r_0|k|=k_{\rm GL}+\sum_{n=1}^{\infty} \epsilon^n k^{(n)},
\end{eqnarray}
where recall that the instability is critical when $r_0|k|=k_{GL}$.  The periodicity of the torus allows us to express the $x$ and $y$ dependence of $q_i^{(n)}$ as a sum of Fourier modes.  The expansion \eqref{expansion} is placed into the Einstein DeTurck equation \eqref{EdeT} and expanded order by order in $\epsilon$.

At $O(\epsilon)$, we have an eigenvalue problem for $\{k_{\rm GL},\, q_i^{(1)} \}$, where all perturbations with the same $r_0|k|=k_\mathrm{GL}$ give the same eigenvalue problem.  We therefore have some freedom to choose the perturbation, which will affect results at higher order. We are only interested in perturbations generated by multiple wavevectors with the same $|k|$. 

For $\gamma\neq1/2$, we take perturbations generated by \eqref{pert1} which takes the form
\begin{equation}\label{O1gammas}
q_{1,2,3}^{(1)}=\frak f_{1,2,3}^{(1)}(z)\left[\cos(x)-\cos(y)\right]\;,
\end{equation} 
while for $\gamma=1/2$, we take a perturbation generated by both \eqref{pert1} and \eqref{pert2}, which goes as
\begin{equation}\label{O1half}
q_{1,2,3}^{(1)}=\frak f_{1,2,3}^{(1)}(z)\left[\cos(x)-\cos(y)+\cos(x+y) \right]\;.
\end{equation} 
Up to symmetries, the perturbations above are actually quite general. We have used translation invariance to fix the phases, and the relative amplitudes (including the amplitude of the $\sin(x+y)$ and $\cos(x+y)$ terms in the $\gamma=1/2$ case) are determined at higher orders in perturbation theory, and we have fixed them retroactively.  The remaining functions at this order vanish due to the gauge condition $\xi^{(1)}_\mu=0$.  

As expected, the equations of motion at linear order for both of these perturbations are identical, and independent of $\gamma$.  They consist of two algebraic relations that can be used to determine $q_{1,3}^{(1)}(z)$ in terms of $q_{2}^{(1)}(z)$ and $q_{2}^{(1)\prime}(z)$, as well as the following eigenvalue problem with $k_{\rm GL}^2$ appearing as an eigenvalue:
\begin{eqnarray}\label{linearorder}
&&\hspace{-0.7cm} q_2^{(1)\prime\prime}(z)+\frac{3+4 z^2-15 z^4}{z-4 z^3+3 z^5}\,q_2^{(1)\prime}(z)\nonumber\\
&& +4 \frac{8 \left(1-z^2\right)^3-k_{\rm GL}^2 \left(1-3 z^2\right)}{\left(1-z^2\right)^4 \left(1-3 z^2\right)} \,q_2^{(1)}(z)=0.
\end{eqnarray}
Together with the boundary conditions that $q_2^{(1)}$ must be regular at the horizon $z=0$ and vanish exponentially at the asymptotic boundary $z=1$, we can solve this problem numerically using methods described in detail in \cite{Dias:2015nua}. We find that $k_{\rm GL} \simeq 0.87616040$.   

For concreteness, we now describe the higher order perturbative analysis for the $\gamma=1/2$ case. The details of the $\gamma\neq1/2$ case can be found in the Appendix.   
At $O(\epsilon^2)$, the Einstein DeTurck equation is sourced by the linear order solution $\{k_{\rm GL},\, q_i^{(1)} \}$.  Then the allowed Fourier modes for the $n=2$ functions come from squaring the linear combination of modes we had at linear order. For example, in the $\gamma=1/2$ case, $q_i^{(2)}$ for $i=1,\ldots,6$ take the form
\begin{eqnarray}\label{2ndorder}
&&\hspace{-0.6cm} q_i^{(2)}(z,x,y)= \frak{f}_i^{(2,0)}(z)+\frak{f}_i^{(2,1)}(z) \cos(x)\\ 
&& +\frak{f}_i^{(2,2)}(z) \cos(y)+\frak{f}_i^{(2,3)}(z) \cos(x+y)\nonumber\\ 
&& +\frak{f}_i^{(2,4)}(z) \cos(x-y)+\frak{f}_i^{(2,5)}(z) \cos(2x)\nonumber\\ 
&& +\frak{f}_i^{(2,6)}(z)\cos(2y)+\frak{f}_i^{(2,7)}(z)\cos[2(x+y)]\nonumber\\ 
&&+ \frak{f}_i^{(2,8)}(z)\cos(2x+y)+\frak{f}_i^{(2,9)}(z) \cos(x+2y),\nonumber
\end{eqnarray}
and for $i=7,8$ one has the same with $\cos \to \sin$ and $ \frak{f}_{7,8}^{(2,0)}(z)=0$. There are a total of 78 functions $\frak{f}_i^{(2,\alpha)}(z)$ which need to determined.  Since Fourier coefficient decouple, each fixed $\alpha$ gives an independent system of differential equations to be solved numerically using the methods detailed in \cite{Dias:2015nua}.  Note that three of these systems with $\alpha=1,2,3$ (\ie the three corresponding to Fourier coefficients that were also excited at linear order) depend upon the correction $k^{(1)}$. The solution to their differential equations will determine both $\frak{f}_i^{(2,\alpha)}(z)$ and $k^{(1)}$ independently.  Since $k^{(1)}$ is unique, it is a consistency check to verify that all three systems yield the same $k^{(1)}$, which they do within machine precision. Numerical convergence is shown in the Appendix. 

The computation continues in a similar manner for higher orders in $\epsilon$. Details can be found in the Appendix. In the end, we stop at $O(\epsilon^4)$ and find
\begin{eqnarray}\label{kcorrections}
k_{\rm GL} &\simeq& 0.87616040 \,, \qquad
k^{(1)} \simeq  -0.14881243 \,,\nonumber\\
k^{(2)} &\simeq & 0.58519439  \,, \qquad  k^{(3)} \simeq  0.47526091\,.
\end{eqnarray}
At $O(\epsilon^4)$, we can compute the first perturbative corrections to various thermodynamic potentials, and we have decided not continue to higher order. 

The calculation for $\gamma\neq1/2$ proceeds in a similar fashion.  Some differences in this case are that $k^{(n)}=0$ when $n$ is odd, and that one needs to reach $O(\epsilon^5)$ in perturbation theory to find thermodynamic corrections.  To obtain our results, the ODEs are solved numerically for a number of specific values of $\gamma$. 

{\bf~Thermodynamics --} We now compute the thermodynamic quantities of the perturbative solutions. The temperature of \eqref{ansatz} is  $T_H=1/(4\pi r_0)$  and the entropy $S_H$ is the horizon area divided by $4 G_N$ ($G_N$ is the six-dimensional Newton's constant). The energy $E$ is computed from the asymptotic decay of the gravitational field using the Hamiltonian formalism presented  in \cite{Harmark:2004ch}. The Helmoltz free energy is then $F=E-T_H S_H$. We compute the dimensionless densities in terms of the torus volume $V_{\mathbb{T}^2}=r_0^2 4\pi^2\overline\lambda^2 (1-\gamma^2)^{-1/2}$:  
\begin{eqnarray}
&\tau_H=T_H V_{\mathbb{T}^2}^{1/2}\;,\qquad\sigma_H=S_H G_N/V_{\mathbb{T}^2}^2\;,\nonumber\\
&{\cal E} =E G_N/V_{\mathbb{T}^2}^{3/2}\;,\qquad \mathit{f}=F G_N/V_{\mathbb{T}^2}^{3/2}\;.
\end{eqnarray}

As we have mentioned earlier, semi-nonuniform black branes in this scenario are never thermodynamically preferred over uniform solutions.  Therefore, we only need to compare our fully nonuniform solutions with the uniform ones. To compare these solutions in the microcanonical ensemble, we take the entropy difference at the same energy $\Delta \sigma_H=\sigma_H({\cal E})-\overline\sigma_H(\cal E)$, where $\sigma_H$ and $\overline\sigma_H$ are the entropy densities of the fully nonuniform, and uniform solutions, respectively.  For comparisons in the canonical ensemble, we similarly define the free energy difference $\Delta f=f(\tau_H)-\overline f(\tau_H)$. 

For lattices with $\gamma=1/2$, to order $\mathcal{O}\left(\epsilon^3\right)$, the thermodynamic densities are
\begin{eqnarray}\label{thermo}
\Delta \sigma_H &\simeq & 0.00183324 \, \epsilon^3\,,\nonumber \\
\tau_H &\simeq &  0.61322672+ 0.10415417\, \epsilon -0.39188874 \, \epsilon^2 \nonumber \\
&& \quad-0.46876238 \, \epsilon^3 \,, \nonumber \\
{\cal E} &\simeq & 0.06488422-0.01102033 \, \epsilon +0.06319328\, \epsilon^2 \nonumber \\
&& \quad+  0.05646878 \, \epsilon^3 \,,\nonumber \\
\Delta \mathit{f} &\simeq & -0.00112418 \, \epsilon^3\,.
\end{eqnarray}

\begin{figure} \centering
\includegraphics[width=.47\textwidth]{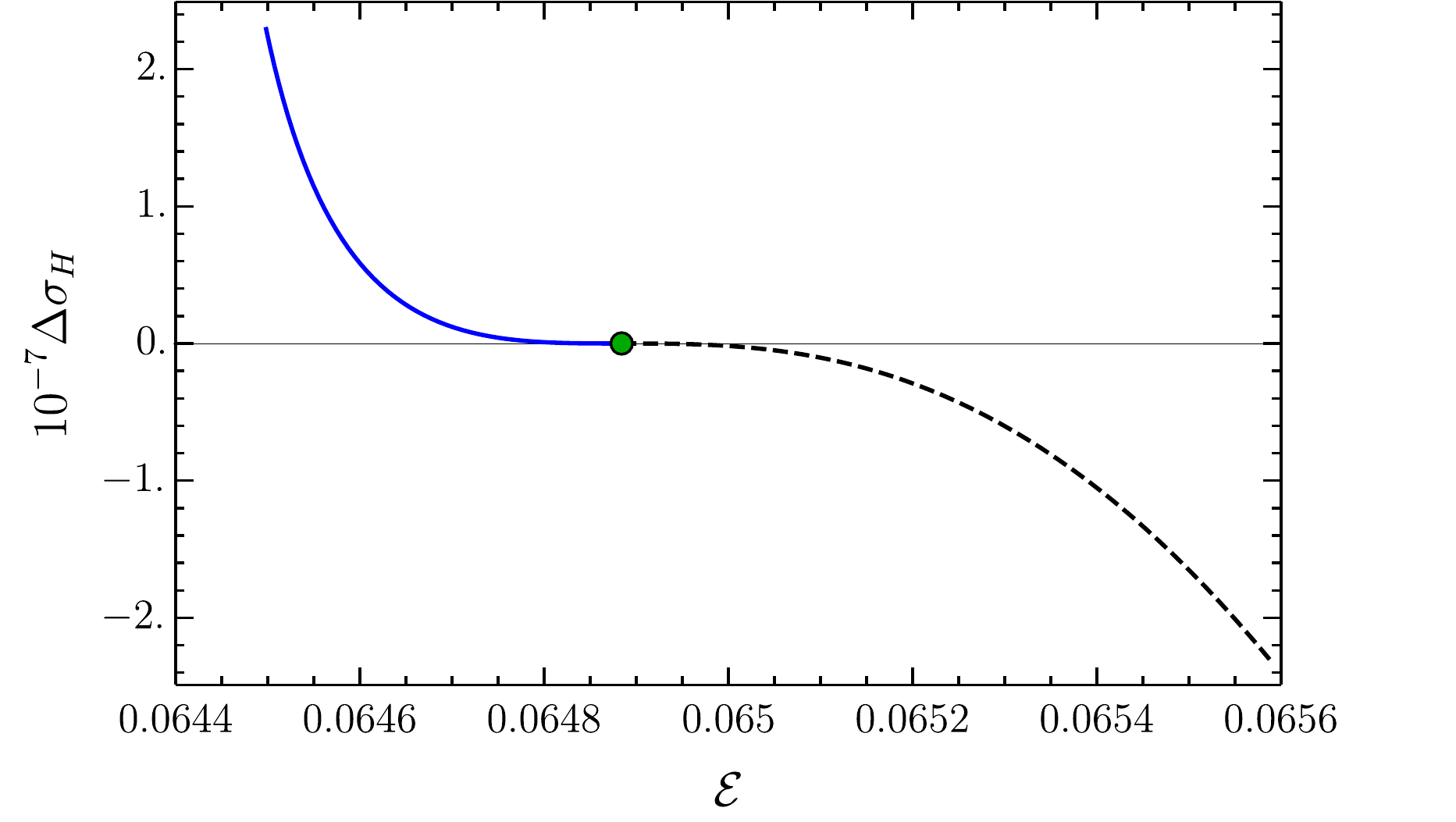}
\caption{Phase diagram in the microcanonical ensemble for  $\gamma=1/2$: entropy density difference $\Delta\sigma_H$ between the triangular (upper blue curve) or hexagonal (lower black curve) nonuniform black branes and the uniform black brane for a given energy density ${\cal E}$. The green dot locates where the uniform black brane is critically unstable, with lower energies being fully unstable.}\label{fig:micro}
\end{figure}

The difference in entropy densities between the $\gamma=1/2$ lattice solutions and the uniform membrane is shown in Fig. \ref{fig:micro}.  The green diamond with ${\cal E}={\cal E}_{\rm GL}=3^{1/4} k_{\rm GL}/(4\sqrt{2}\pi)\simeq 0.06488422$ is the critical point of the instability where uniform branes are unstable at smaller energies.  From this plot, we see that fully nonuniform branes have higher entropy where the uniform solutions are unstable, and are thus are a plausible endpoint to the instability. 

The phase diagram of $\gamma=1/2$ in the canonical ensemble can be found in the Appendix. The main conclusion is that solutions that are dominant in the microcanonical ensemble are also dominant in the canonical ensemble.

The curve in Fig. \ref{fig:micro} extends to both higher and lower energies, which is related to the fact that $\Delta \sigma_H\sim\epsilon^3$ in \eqref{thermo}. The fact that this power is odd rather than even is equivalent to the fact that $\delta g$ and $-\delta g$ are physically distinct perturbations. This distinction can be interpreted as the difference between triangular and hexagonal arrangements. We contrast this with the nonuniform strings, where translation invariance implies that $\delta g$ and $-\delta g$ are physically equivalent \cite{Gubser:2001ac,Wiseman:2002zc,Sorkin:2004qq}. Nonuniform strings therefore only extend towards higher or lower energies (not both), depending on the dimension. 

\begin{figure} \centering
\includegraphics[width=.47\textwidth]{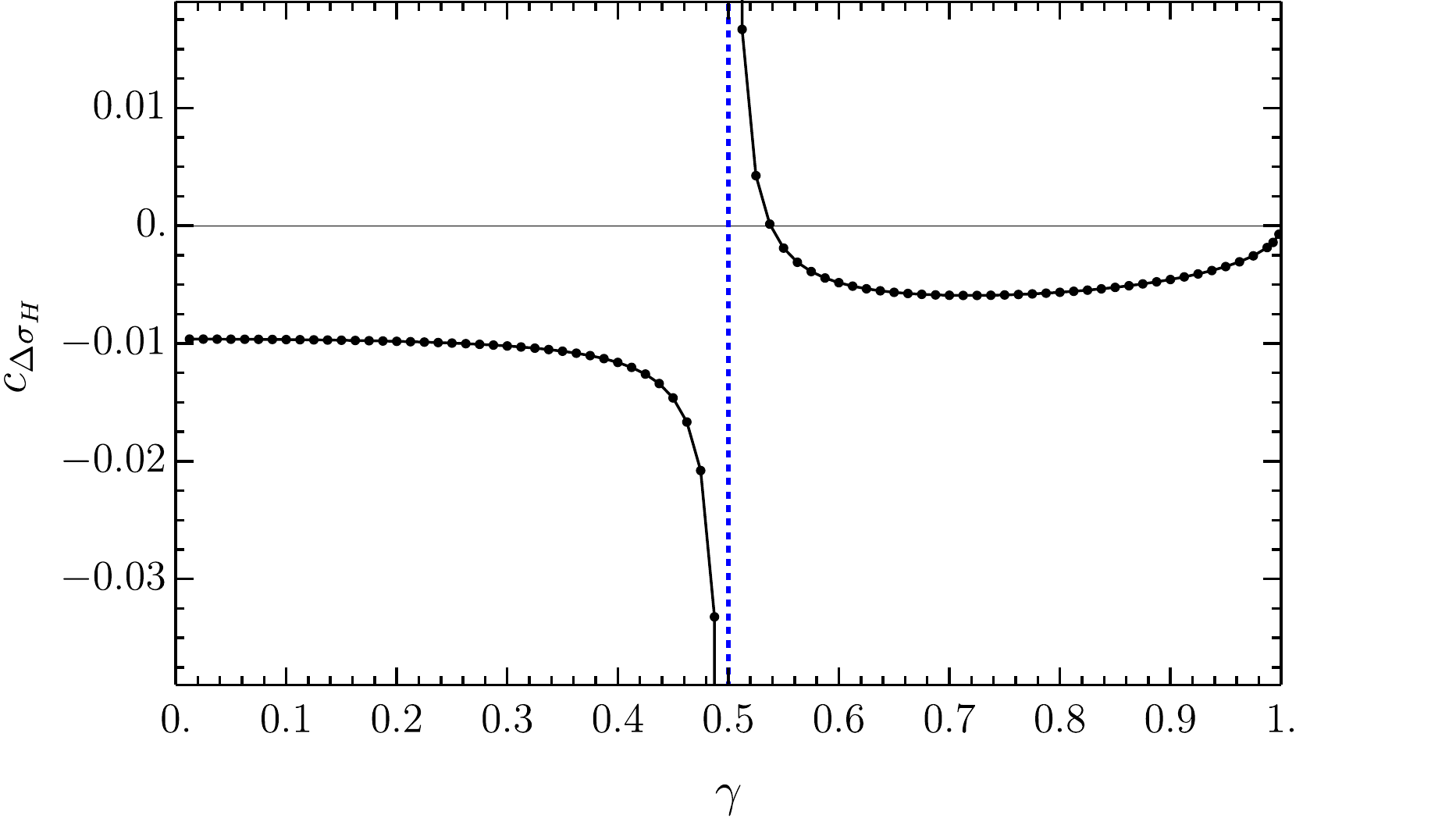}
\caption{Coefficient $c_{\Delta \sigma_H}$ of the entropy density difference $ \Delta \sigma_H\sim c_{\Delta \sigma_H} \epsilon^4$ as a function of the lattice angle $\gamma=\sin\theta$ for $\gamma\neq1/2$. Perturbation theory breaks down at $\gamma=1/2$ and the radius of convergence of our solutions decreases as this critical value is approached.}\label{fig:sigmagamma}
\end{figure}

Consider now the  $\gamma\neq 1/2$ solutions. In these cases, the leading correction to the entropy and free energy differences go as 
\begin{equation}\label{gammanothalfthermo}
\Delta \sigma_H= c_{\Delta \sigma_H} \epsilon^4 +\ldots\;,\qquad \Delta  \mathit{f}= c_{\Delta\mathit{f}} \epsilon^4 + \ldots\;,
\end{equation}
where the coefficients $c_{\Delta \sigma_H}$ and  $c_{\Delta\mathit{f}}$ depend on $\gamma$. Note that we conclude that the even power of $\epsilon$ in \eqref{gammanothalfthermo} implies that each $\gamma$ yields a single branch of lattice solutions (not two, like the $\gamma=1/2$ case). In Fig. \ref{fig:sigmagamma} we plot the coefficient $c_{\Delta \sigma_H}$ as a function of $\gamma$. Observe that $c_{\Delta \sigma_H}<0$ for most values of $\gamma$. Therefore, fully nonuniform black branes with those values of $\gamma$ do not dominate the microcanonical ensemble near the zero mode. However, there is a small window $1/2<\gamma\lesssim0.538$ where $c_{\Delta \sigma_H}>0$. In this window, the fully nonuniform branes are preferred over the uniform phase. Moreover, we also find that within this window, the fully nonuniform brane solutions extend from the zero mode towards lower energies $\mathcal E<\mathcal E_{\mathrm {GL}}$ where the uniform solutions are unstable, but otherwise extends towards higher energies. 

{\bf~Discussion --}  To summarise our results thus far, we have found black branes in $\mathbb{R}^{1,3}\times\mathbb{T}^2$ that are fully inhomogeneous in the torus directions by perturbing about the zero mode.  These separate into the case where the torus contains triangular/hexagonal symmetry $\gamma=1/2$, and otherwise $\gamma\neq1/2$. In the $\gamma=1/2$ case, these nonuniform solutions depend upon the sign of the perturbation, and in the $\gamma\neq1/2$ case they do not. Surprisingly, one branch of the $\gamma=1/2$ branes and some branes with $\gamma>1/2$ are thermodynamically preferred over both the uniform brane and semi-nonuniform black branes in both microcanonical and canonical ensembles. Due to these thermodynamic considerations, it is natural to conjecture that an unstable uniform black membrane with ${\cal E}<{\cal E}_{\rm GL}$ and $\tau_H>\tau_{\rm GL}$ will evolve in time towards a fully-nonuniform black membrane. 

However, localised solutions are expected to exist that might compete with some of these nonuniform solutions. In the space of static solutions, nonuniform black strings are known to eventually connect to localised black holes (which have spatial horizon topology $S^3$) through a topology-changing conical transition \cite{Kudoh:2004hs}. Likewise in our case, we expect spherical $S^4$ black holes to exist.  However, a direct transition from fully nonuniform black branes to localised black holes requires a transition where an entire circle pinches off on the horizon simultaneously in moduli space.  We find such a scenario unlikely due to the lack of symmetry.  Instead, it is more plausible that the horizon pinches off on specific points, leading to mesh-shaped black objects resembling connected black strings. Those might later proceed through a second topology transition towards localised black holes. Sorting out this phase diagram requires constructing these hypothetical localised solutions, as well as extending our results for nonuniform branes to nonperturbative regions. 

\begin{figure} \centering
\includegraphics[width=.85\textwidth]{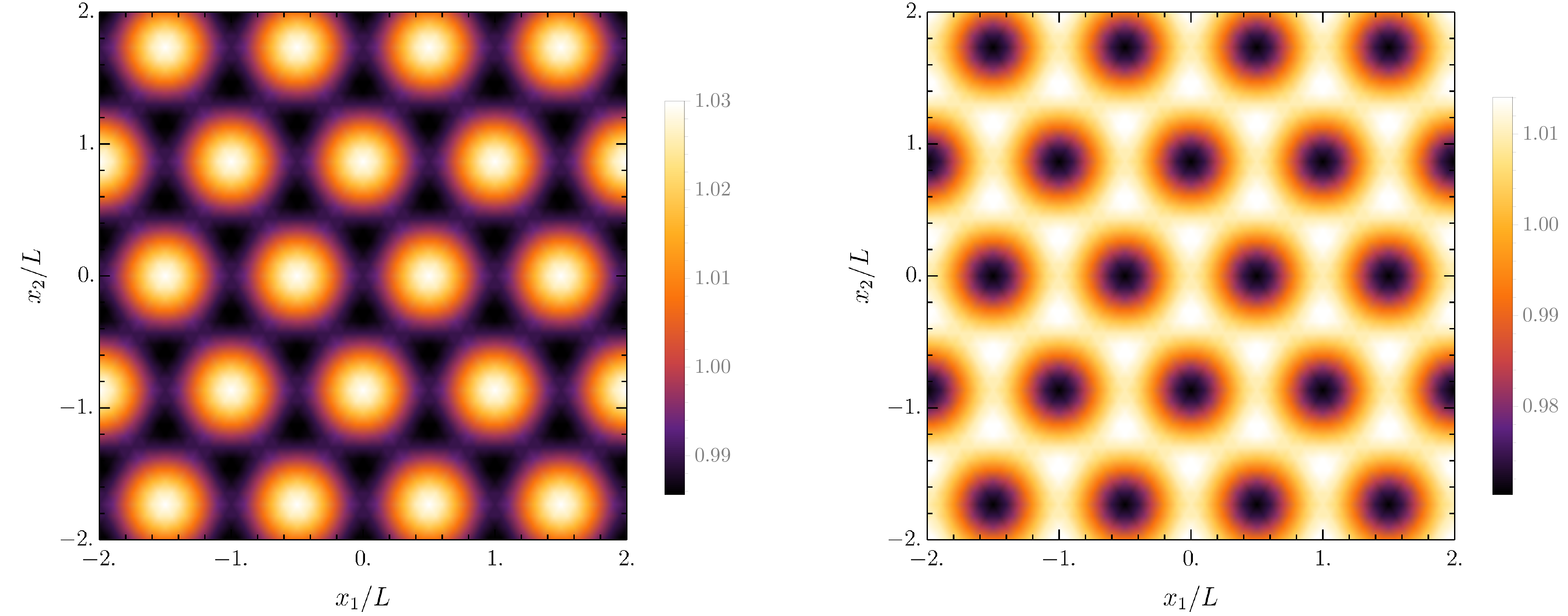}
\caption{Structure of triangular and hexagonal nonuniform black branes ($\gamma=1/2$, $\epsilon=\pm10^{-2}$).
Radius of the $S^2$ at the horizon (which is proportional to the square root of the entropy density) as a function of the lattice directions $x_1$ and $x_2$ for the black brane with triangular (left) and hexagonal (right) arrangements. The lighter yellow regions represent the locations  of the lattice points where the entropy density is higher.
}\label{fig:latticeRS2}
\end{figure}

Note that the extra symmetries of the $\gamma=1/2$ case allow for two arrangements of localised black holes that our fully nonuniform black branes may eventually connect to. One of these is a hexagonal arrangement, and the other is a triangular arrangement. The $S^2$ radius of the black brane is shown in Fig. \ref{fig:latticeRS2}, where it is possible to extrapolate which arrangement of localised black holes our fully nonuniform black branes are approaching. We see that the higher-entropy branch of these black branes approaches the triangular arrangement (left panel), while the lower-entropy branch approaches a hexagonal arrangement (right panel). Because a hexagonal arrangement of localised black holes would contain two black holes per torus volume, we expect this situation to be unstable. 

It is worth considering what happens to these solutions at higher dimensions, where the Schwarzschild black hole is replaced by Schwarzschild-Tangherlini. In $D\geq 14$ dimensions, semi-nonuniform black branes will become preferred over uniform black branes, and it would be interesting to see how they compete with fully nonuniform branes. Note that for $\gamma>1/2$, there is an additional semi-nonuniform solution coming from the perturbations \eqref{pert2} since they have longer wavelength than those of \eqref{pert1} that generated the fully nonuniform solutions. 

We also mention that our methods here are not specific to two brane directions, though the situation grows in complexity with increasing brane dimension. It is interesting to note that for two brane directions, the most entropic nonuniform phase moves towards the triangular arrangement of black holes, which is also the densest sphere packing in two dimensions.  It is therefore tempting to propose a connection between thermodynamically preferred Kaluza-Klein black holes and the mathematical problem of sphere packing.  However, we expect the preferred localised black hole phase to contain only one black hole per torus volume.  If there is indeed a connection to sphere packing, it would be the problem of packing spheres on fundamental lattice sites, and not the more sophisticated general sphere packing problem.

Our work raises interesting questions for the study of unstable gravitational systems with multiple extended directions. Although our study was restricted to asymptotically Kaluza-Klein flat spacetimes, we expect similar physics to be present in the context of asymptotically locally anti-de Sitter spacetimes. For instance, the analysis of the so called \emph{spinoidal instability} performed in \cite{Attems:2017ezz} and \cite{Janik:2017ykj} was restricted to co-homogeneity $2+1$, and our work raises the possibility of finding interesting new phases with fewer symmetries. It would also be interesting to pursue the phase structure of the holographic duals of $p-$branes compactified on $\mathbb{T}^d$ torus which was initiated in \cite{Hanada:2007wn}.

{\bf~Acknowledgements --} 
O.J.C.D. is supported by the STFC Ernest Rutherford grants ST/K005391/1 and ST/M004147/1. O.J.C.D. further acknowledges support from the STFC ``Particle Physics Grants Panel (PPGP) 2016", grant ref. ST/P000711/1. J.E.S. was supported in part by STFC grants PHY-1504541 and ST/P000681/1.   B.W. is supported by NSERC.

\onecolumngrid

\twocolumngrid
\newpage
\onecolumngrid  \vspace{1cm} 
\appendix 
\begin{center}  
{\Large\bf Appendix} 
\end{center} 

\section{Details of the perturbation theory for $\gamma=1/2$}

Here, we provide details to our perturbative calculation for the $\gamma=1/2$ lattices (\ie triangular/hexagonal lattices). Our method is similar to the one used by \cite{Gubser:2001ac,Wiseman:2002zc,Sorkin:2004qq} to explore the existence of asymptotically $\mathbb{R}^{1,4}\times S^1$ non-uniform black strings. 

We perturb the metric functions in a power series written as
\begin{eqnarray}\label{expansion2}
q_i(z,x,y)=\bar{q}_{i}+\sum_{n=1}^{\infty} \epsilon^n \,q_i^{(n)}(z,x,y), \qquad r_0|k|=k_{\rm GL}+\sum_{n=1}^{\infty} \epsilon^n k^{(n)},
\end{eqnarray}
where $\bar{q}_{i}$ describes the uniform black membrane.  The torus implies that the perturbation functions $q_i^{(n)}$ can be expanded in a Fourier series.

But triangular/hexagonal lattice symmetries (as shown in Fig. \ref{fig:latticeRS2}) limit the number of Fourier coefficients that are available. As explained in the discussion of of \eqref{pert1}-\eqref{pert2} of the the main text, for $\gamma=1/2$ we have three Fourier modes with the same wavevector norm: $\cos(x), \cos(y)$ and $\cos(x+y)$.   The equations of motion and boundary conditions at $O(\epsilon)$ imply
\begin{eqnarray}\label{linearorderSolution}
k_{\rm GL}&\simeq& 0.87616040\,; 
\nonumber\\
q_{1}^{(1)}(z,x,y)&=&\left( \frac{5 z^2+1}{1-3 z^2}\,q_{2}^{(1)}(z)+\frac{z \left(1-z^2\right)}{1-3 z^2}\,q_{2}^{(1)\prime}(z)  \right) \big[\cos(x)-\cos(y)+\cos(x+y) \big],
\nonumber\\
q_{2}^{(1)}(z,x,y)&=&q_{2}^{(1)}(z) \big[\cos(x)-\cos(y)+\cos(x+y) \big]
\,, 
\nonumber\\
q_{3}^{(1)}(z,x,y)&=& -\left( \frac{z^2+1}{1-3 z^2} \,q_{2}^{(1)}(z)+ \frac{z \left(1-z^2\right)}{2 \left(1-3 z^2\right)} \,q_{2}^{(1)\prime}(z)  \right) \big[\cos(x)-\cos(y)+\cos(x+y) \big],
\nonumber\\
q_{i}^{(1)}(z,x,y)&=& 0\,, \quad\hbox{for}\:\: i=4,\cdots,8\,;
\end{eqnarray}
where $k_{\rm GL}$ and $q_{2}^{(1)}(z)$ are the (numerical) solutions of the eigenvalue equation
\begin{equation}\label{linearorder2}
q_2^{(1)\prime\prime}(z)+\frac{3+4 z^2-15 z^4}{z-4 z^3+3 z^5}\,q_2^{(1)\prime}(z) +4 \frac{8 \left(1-z^2\right)^3-k_{\rm GL}^2 \left(1-3 z^2\right)}{\left(1-z^2\right)^4 \left(1-3 z^2\right)} \,q_2^{(1)}(z)=0.
\end{equation}

The linear order solution \eqref{linearorderSolution}-\eqref{linearorder2} now sources the second order solution at $O(\epsilon^2)$. The Fourier modes that are excited at this order come from squaring the sum of the three $O(\epsilon)$ modes.  In full, they can be written
\begin{eqnarray}\label{2ndorder2}
 q_i^{(2)}(z,x,y)&=& \frak{f}_i^{(2,0)}(z)+\frak{f}_i^{(2,1)}(z) \cos(x) +\frak{f}_i^{(2,2)}(z) \cos(y)+\frak{f}_i^{(2,3)}(z) \cos(x+y) \nonumber\\ 
&& +\frak{f}_i^{(2,4)}(z) \cos(x-y) +\frak{f}_i^{(2,5)}(z) \cos(2x) +\frak{f}_i^{(2,6)}(z)\cos(2 y)+\frak{f}_i^{(2,7)}(z)\cos[2(x+y)]\nonumber\\ 
&&+ \frak{f}_i^{(2,8)}(z)\cos(2x+y)+\frak{f}_i^{(2,9)}(z) \cos(x+2y), \quad  \quad \hbox{for} \:\: i=1,\cdots,6\,,
\nonumber\\ 
 q_i^{(2)}(z,x,y)&=& 0+\frak{f}_i^{(2,1)}(z) \sin(x) +\frak{f}_i^{(2,2)}(z) \sin(y)+\frak{f}_i^{(2,3)}(z) \sin(x+y) \nonumber\\ 
&& +\frak{f}_i^{(2,4)}(z) \sin(x-y) +\frak{f}_i^{(2,5)}(z) \sin(2 x) +\frak{f}_i^{(2,6)}(z)\sin(2y)+\frak{f}_i^{(2,7)}(z)\sin[2(x+y)]\nonumber\\ 
&&+ \frak{f}_i^{(2,8)}(z)\sin(2x+y)+\frak{f}_i^{(2,9)}(z) \sin(x+2y), \quad  \quad \hbox{for} \:\: i=7,8\,,
\end{eqnarray}
We wish to determine the various $ \frak{f}_i^{(2,\alpha)}(z)$, with $\alpha=0,\cdots 9$ (a total of 78 functions since $\frak{f}_{7,8}^{(2,0)}(z)=0$). Since each Fourier mode $\alpha$ decouples from the others, we solve an independent coupled ODE system for each $\alpha$ (a second-order system of 6 ODEs for $\alpha=0$ and 8 for the others). These ODEs here, and at higher orders in $\epsilon$ take the form
\begin{equation}\label{ODEprototype}
\mathcal{O}_{ij} q_j^{(n,\alpha)}=S_i^{(n,\alpha)}+k^{(n-1)} s_i^{(n,\alpha)}
\end{equation} 
 where $\mathcal{O}_{ij}$ is a second order differential operator and  $S_i^{(n,\alpha)},s_i^{(n,\alpha)}$ are sourced by the lowest order solution \eqref{linearorderSolution}-\eqref{linearorder2}. For $\alpha= 1,2,3$, \ie precisely the modes that were excited at linear order, one has $s_i^{(n,\alpha)}\neq 0$ so we can use each of these systems to determine $k^{(1)}$. We verify for consistency that we obtain the same $k^{(1)}$ $-$ namely the one given in \eqref{kcorrections} $-$ from each $\alpha= 1,2,3$.
 
The calculation at higher orders follows a similar process. At $O(\epsilon^3)$, we have a total of 19 Fourier modes, each of the form $\cos({\cal A}^{(3,\alpha)})$, where 
\begin{equation}
{\cal A}^{(3,\alpha)}\in\{ 0,x,y,x+y,x-y,2 x,2 y,2 (x+y),2 x+y,x+2 y,
3 x,3 y,3 (x+y),2 x-y,x-2 y,3 x+y,x+3 y,3 x+2 y,2 x+3 y\}\;,
\end{equation}
which are precisely those that follow from taking the cubic power of the sum of the three linear-order modes.  The perturbation functions therefore take the form
\begin{eqnarray}\label{3rdorder}
 q_i^{(3)}(z,x,y)&=& \frak{f}_i^{(3,0)}(z)+\frak{f}_i^{(3,1)}(z) \cos(x) +\frak{f}_i^{(3,2)}(z) \cos(y)+\frak{f}_i^{(3,3)}(z) \cos(x+y) \nonumber\\ 
&& +\frak{f}_i^{(3,4)}(z) \cos(x-y) +\frak{f}_i^{(3,5)}(z) \cos(2x) +\frak{f}_i^{(3,6)}(z)\cos(2y)+\frak{f}_i^{(3,7)}(z)\cos[2(x+y)]\nonumber\\ 
&&+ \frak{f}_i^{(3,8)}(z)\cos(2x+y)+\frak{f}_i^{(3,9)}(z) \cos(x+2y)
+\frak{f}_i^{(3,10)}(z) \cos(3 x)+\frak{f}_i^{(3,11)}(z) \cos(3 y)\nonumber\\ 
&& +\frak{f}_i^{(3,12)}(z) \cos[3(x+y)]+\frak{f}_i^{(3,13)}(z) \cos(2 x-y)+\frak{f}_i^{(3,14)}(z) \cos(x-2 y)
\nonumber\\ 
&&+\frak{f}_i^{(3,15)}(z) \cos(3 x+y)+\frak{f}_i^{(3,16)}(z) \cos(x+3 y)+\frak{f}_i^{(3,17)}(z) \cos(3 x+2)\nonumber\\ 
&&+\frak{f}_i^{(3,18)}(z) \cos(2 x+3 y)
, \quad  \quad \hbox{for} \:\: i=1,\cdots,6\,,
\nonumber\\ 
 q_i^{(3)}(z,x,y)&=&0+\frak{f}_i^{(3,1)}(z) \sin(x) +\frak{f}_i^{(3,2)}(z) \sin(y)+\frak{f}_i^{(3,3)}(z) \sin(x+y) \nonumber\\ 
&& +\frak{f}_i^{(3,4)}(z) \sin(x-y) +\frak{f}_i^{(3,5)}(z) \sin(2 x) +\frak{f}_i^{(3,6)}(z)\sin(2 y)+\frak{f}_i^{(3,7)}(z)\sin[2(x+y)]\nonumber\\ 
&&+ \frak{f}_i^{(3,8)}(z)\sin(2x+y)+\frak{f}_i^{(3,9)}(z) \sin(x+2y)
+\frak{f}_i^{(3,10)}(z) \sin(6\pi x)+\frak{f}_i^{(3,11)}(z) \sin(3y)\nonumber\\ 
&& +\frak{f}_i^{(3,12)}(z) \sin[3(x+y)]+\frak{f}_i^{(3,13)}(z) \sin(2 x-y)+\frak{f}_i^{(3,14)}(z) \sin(x-2 y)
\nonumber\\ 
&&+\frak{f}_i^{(3,15)}(z) \sin(3 x+y)+\frak{f}_i^{(3,16)}(z) \sin(x+3 y)+\frak{f}_i^{(3,17)}(z) \sin(3 x+2)\nonumber\\ 
&&+\frak{f}_i^{(3,18)}(z) \cos(2 x+3 y), \quad  \quad \hbox{for} \:\: i=7,8\,,
\end{eqnarray}
We now have 150 functions $\frak{f}_i^{(3,\alpha)}(z)$ to solve for. As before, each Fourier mode $\alpha$ decouples from all others, so we have 19 independent systems of ODEs to solve (one for each $\alpha$). These ODEs again take the form \eqref{ODEprototype}. Again, the systems corresponding to $\alpha=1,2,3$, and only these, have $s_i^{(3,\alpha)}\neq 0$, and hence depend on $k^{(2)}$. Therefore, for each of these three $\alpha$'s we solve the equations of motion to find $\frak{f}_i^{(3,\alpha)}(z)$ and $k^{(2)}$. These three systems of equations are independent but $k^{(2)}$ is unique, so we must to get the same $k^{(2)}$ in each of them. This $k^{(2)}$ is presented in \eqref{kcorrections}.

To find the wavevector correction $k^{(3)}$, which is required to compute the thermodynamic quantities up to $O(\epsilon^3)$, we still need to find the metric solutions at order $O(\epsilon^4)$ since $k^{(3)}$ only appears in the equations of motion at this order (in practice it is enough to analyse the $\cos x, \cos y$ and $\cos (x+y)$ Fourier sectors since $k^{(3)}$ only appears in the equations associated to these modes). At this order, a total of 31 Fourier modes  are excited. Schematically, each of these is of form $\cos({\cal A}^{(4,\alpha)})$, with 
\begin{eqnarray}
{\cal A}^{(4,\alpha)}&\in&\{0,x,2 x,3 x,4 x,y,2 y,3 y,4 y,x\pm y,2 (x\pm y),3 (x+y),4 (x+y),x\pm 2 y,2 x\pm y,2 (x+2 y),2 (2 x+y),\nonumber\\
&& \:\:\:x\pm 3 y,3 x\pm y,4 x+y,x+4 y,3 x+2 y,2 x+3 y,4 x+3 y,3 x+4 y\}, 
\end{eqnarray}
which are precisely those that follow from taking the fourth power of the sum of the three linear-order modes in \eqref{linearorderSolution}. Solving the corresponding ODE system we determine the associated Fourier coefficients $\frak{f}_i^{(4,\alpha)}(z)$ for $\alpha=0,\cdots,30$ and $k^{(3)}\neq 0$ given in \eqref{kcorrections}.

We stop our analysis at this order $O(\epsilon^4)$,  since this is the first order where the leading contribution to the entropy difference $\Delta \sigma_H$, and free energy difference $\Delta  \mathit{f}$ are obtained. Namely, the entropy, temperature, energy and free energy densities of the $\gamma=1/2$ black membrane solutions are given by  
\begin{eqnarray}\label{thermo2}
\sigma_H &\simeq &0.05290394 -0.01797106\,\epsilon +0.10457659\, \epsilon^2 +0.07641519 \, \epsilon^3\,,\nonumber \\
\tau_H &\simeq &  0.61322672+ 0.10415417\, \epsilon -0.39188874 \, \epsilon^2 -0.46876238 \, \epsilon^3 \,, \nonumber \\
{\cal E} &\simeq & 0.06488422-0.01102033 \, \epsilon +0.06319328\, \epsilon^2  +  0.05646878 \, \epsilon^3 \,,\nonumber \\
 \mathit{f} &\simeq & 0.03244211-0.00551017\,\epsilon +0.02166834 \, \epsilon^2 +0.01647358 \, \epsilon^3\,.
\end{eqnarray}
To determine which phase is thermodynamically preferred in a given ensemble, we must compare \eqref{thermo2} with those for the uniform black membrane. In the microcanonical ensemble, one keeps the energy density fixed and the dominant solution is the one with higher entropy density. In the canonical ensemble, one keeps the dimensionless temperature fixed and the dominant solution is the one with lower free energy density. For this analysis, it is thus important to write the entropy density of the uniform black brane as a function of its energy density, and its free energy density as a function of its dimensionless temperature:
\begin{equation}\label{thermoUnif}
\overline\sigma_H({\cal E})\big|_{\rm unif}=4\pi {\cal E}^2\,, \qquad \overline{f}(\tau_H)\big|_{\rm unif}=\frac{1}{16\pi \,\tau_H}\,.
\end{equation}

For the microcanonical analysis, we can now compute the entropy density difference $\Delta \sigma_H=\sigma_H({\cal E})-\overline\sigma_H(\cal E)$ between nonuniform and uniform black branes with the same energy density ${\cal E}$ to get the result $\Delta \sigma_H \simeq 0.00183324 \, \epsilon^3$ presented in the main text. Similarly for the canonical ensemble, we can  compute the free energy density difference $\Delta f=f(\tau_H)-\overline f(\tau_H)$ at the same dimensionless temperature $\tau_H$ to get the result $\Delta \mathit{f} \simeq  -0.00112418 \, \epsilon^3$ presented in the main text.
 
\begin{figure} \centering
\includegraphics[width=.47\textwidth]{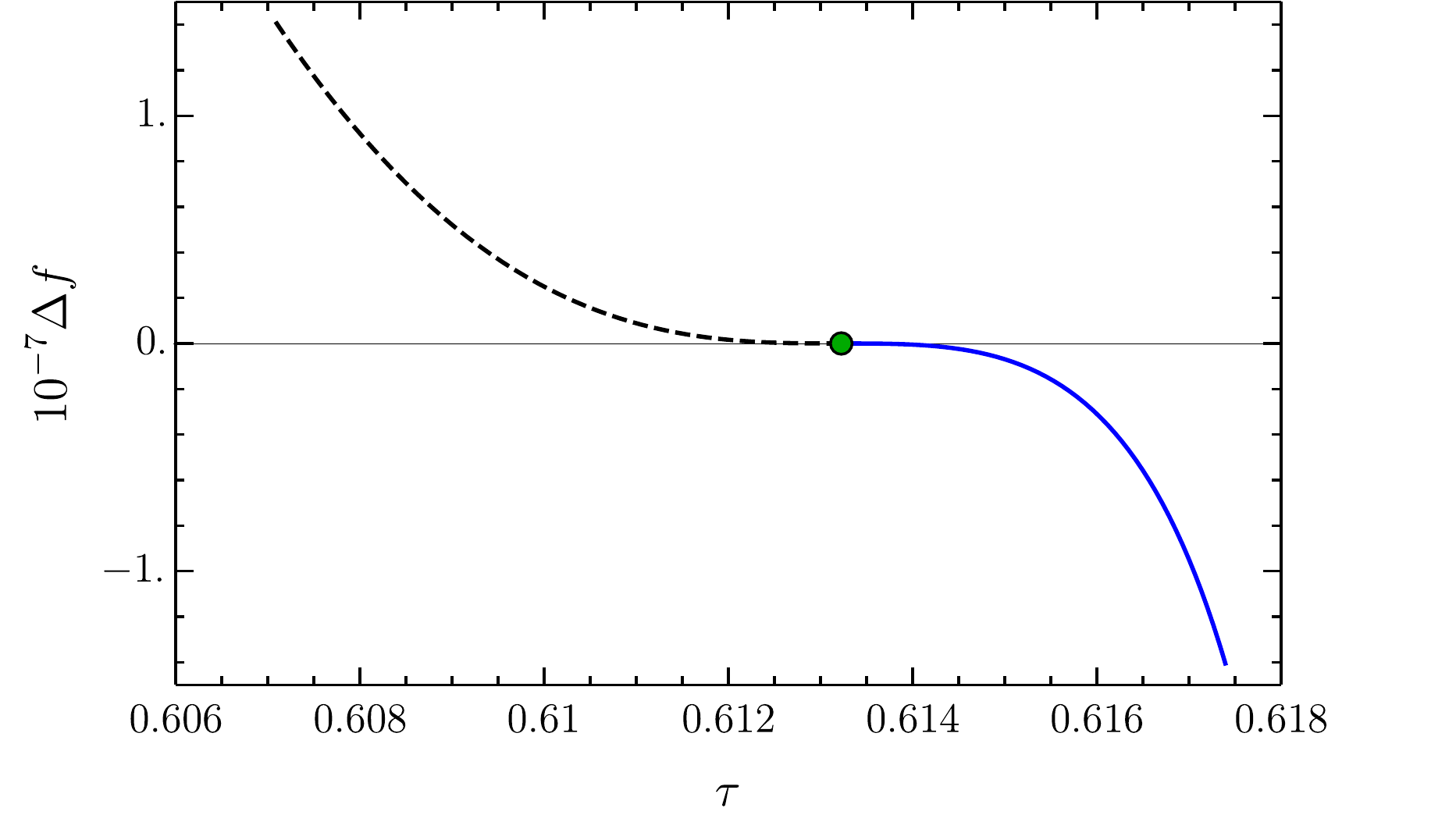}
\caption{Phase diagram in the canonical ensemble: dimensionless free energy density difference $\Delta \mathit{f} $ between the triangular (lower blue curve) and hexagonal (upper dashed black curve) nonuniform black membranes and the uniform black membrane for a given dimensionless temperature $\tau_H$.  Solutions with the lower free energy are preferred. The green dot locates the zero-mode of the instability of uniform branes, with higher temperatures being unstable.}\label{fig:canonical}
\end{figure}

The microcanonical phase diagram was presented in the main text. For completeness, here we show the canonical phase diagram  $\Delta  \mathit{f}$ vs $\tau_H$ in Fig. \ref{fig:canonical}. 

\section{Details of the perturbation theory for $\gamma\neq 1/2$\label{app:gammas}}
Just as we did for $\gamma=1/2$, in the $\gamma\neq1/2$ case, we perturb the metric functions in a power series as described in \eqref{expansion2}. The lattice symmetries limit the number of Fourier coefficients that are excited. For reason explained in the discussion of \eqref{pert1}-\eqref{pert2} of the the main text, these excited modes differ from those present in the $\gamma=1/2$ case studied in the previous section of the appendix. We have know only two Fourier modes with the same wavevector norm: $\cos(x)$ and $\cos(y)$. The equations of motion and boundary conditions at $O(\epsilon)$ imply
\begin{eqnarray}\label{linearorderSolutionB}
k_{\rm GL}&\simeq& 0.87616040\,; 
\nonumber\\
q_{1}^{(1)}(z,x,y)&=&\left( \frac{5 z^2+1}{1-3 z^2}\,q_{2}^{(1)}(z)+\frac{z \left(1-z^2\right)}{1-3 z^2}\,q_{2}^{(1)\prime}(z)  \right) \big[\cos(x)-\cos(y) \big],
\nonumber\\
q_{2}^{(1)}(z,x,y)&=&q_{2}^{(1)}(z) \big[\cos(x)-\cos(y) \big]
\,, 
\nonumber\\
q_{3}^{(1)}(z,x,y)&=& -\left( \frac{z^2+1}{1-3 z^2} \,q_{2}^{(1)}(z)+ \frac{z \left(1-z^2\right)}{2 \left(1-3 z^2\right)} \,q_{2}^{(1)\prime}(z)  \right) \big[\cos(x)-\cos(y) \big],
\nonumber\\
q_{i}^{(1)}(z,x,y)&=& 0\,, \quad\hbox{for}\:\: i=4,\cdots,8\,;
\end{eqnarray}
where $k_{\rm GL}$ and $q_{2}^{(1)}(z)$ are the (numerical) solutions of the eigenvalue equation \eqref{linearorder2}. Note that with respect to the $\gamma=1/2$ case, $q_{2}^{(1)}(z)$ satisfies the same eigenvalue equation \eqref{linearorder2} and   $q_{i}^{(1)}(z)$ and $k_{\rm GL}$ are the same for all $\gamma$'s.

The linear order solution of  \eqref{linearorder2} and \eqref{linearorderSolutionB} now sources the solution at $O(\epsilon^2)$. The excited Fourier modes at this order come from squaring the sum of the two $O(\epsilon)$ modes.  In total, they can be written as
\begin{eqnarray}\label{2ndorder2B}
 q_i^{(2)}(z,x,y)&=& \frak{f}_i^{(2,0)}(z)+\frak{f}_i^{(2,1)}(z) \cos(2x) +\frak{f}_i^{(2,2)}(z) \cos(2y)\nonumber \\ 
 && +\frak{f}_i^{(2,3)}(z) \cos(x+y)+\frak{f}_i^{(2,4)}(z) \cos(x-y),  \quad \quad \hbox{for} \:\: i=1,\cdots,6\,,
\nonumber \\ 
 q_i^{(2)}(z,x,y)&=& 0+\frak{f}_i^{(2,1)}(z) \sin(2x) +\frak{f}_i^{(2,2)}(z) \sin(2y)\nonumber \\ 
 && +\frak{f}_i^{(2,3)}(z) \sin(x+y) +\frak{f}_i^{(2,4)}(z) \sin(x-y), \quad  \quad \hbox{for} \:\: i=7,8\,.
\end{eqnarray}
We need to determine the several $ \frak{f}_i^{(2,\alpha)}(z)$, with $\alpha=0,\cdots 4$ (a total of 38 functions since $\frak{f}_{7,8}^{(2,0)}(z)=0$), for each $\gamma\neq 1/2$. Each Fourier mode $\alpha$ decouples from the others, so the equations reduce to an independent coupled ODE system for each $\alpha$ (a second-order system of 6 ODEs for $\alpha=0$ and 8 for the others) and $\gamma$. These ODEs, as well as those at higher orders in $\epsilon$, take the form \eqref{ODEprototype} where $\mathcal{O}_{ij}$ is a second order differential operator and  $S_i^{(n,\alpha)},s_i^{(n,\alpha)}$ are sourced by the lowest order solution \eqref{linearorderSolutionB}. 

At the current order, $n=2$, one has $s_i^{(n,\alpha)}= 0$ for all the Fourier families (\ie for the five $\alpha$'s). In particular, this implies that $k^{(1)}=0$ (which was not the case for $\gamma=1/2$).   
 To be more precise, note that we can add the Fourier modes $\cos x$ and $\cos y$  to \eqref{2ndorder2B}. However, since these modes are not sourced by the two $\mathcal{O}(\epsilon)$ modes, we find that the associated ODE system is of the form \eqref{ODEprototype} with $S_i^{(n,\alpha)}=0$. There is nevertheless a source term proportional to $k^{(1)}$, \ie $s_i^{(n,\alpha)}\neq 0$, which results from the $k$-expansion in \eqref{expansion2}. However,  the system only has the trivial solution $k^{(1)}=0$ and vanishing  eigenfunctions. The structure of the problem is such that $k^{(n)}=0$ for odd order $n$.
 
The calculation at higher orders follows a similar process. At $O(\epsilon^3)$, we have a total of 8 Fourier modes, each of the form $\cos({\cal A}^{(3,\alpha)})$, where 
\begin{equation}
{\cal A}^{(3,\alpha)}\in\{ x,y,3 x,3 y,2 x+y,x+2 y,2 x-y,x-2 y\},
\end{equation}
which are precisely those that follow from taking the cubic power of the sum of the two linear-order modes in \eqref{linearorderSolutionB}. Namely, the perturbation functions take the form
\begin{eqnarray}\label{3rdorderB}
 q_i^{(3)}(z,x,y)&=& \frak{f}_i^{(3,1)}(z) \cos(x) +\frak{f}_i^{(3,2)}(z) \cos(y)
 +\frak{f}_i^{(3,3)}(z) \cos(3 x)  +\frak{f}_i^{(3,4)}(z) \cos(3 y)  + \frak{f}_i^{(3,5)}(z)\cos(2x+y)
\nonumber\\ 
&& +\frak{f}_i^{(3,6)}(z) \cos(x+2y) +\frak{f}_i^{(3,7)}(z) \cos(2 x-y)+\frak{f}_i^{(3,8)}(z) \cos(x-2 y), \quad  \quad \hbox{for} \:\: i=1,\cdots,6\,,
\nonumber\\ 
 q_i^{(3)}(z,x,y)&=&\frak{f}_i^{(3,1)}(z) \sin(x) +\frak{f}_i^{(3,2)}(z) \sin(y)
 +\frak{f}_i^{(3,3)}(z) \sin(3 x)  +\frak{f}_i^{(3,4)}(z) \sin(3 y)  + \frak{f}_i^{(3,5)}(z)\sin(2x+y)
\nonumber\\ 
&& +\frak{f}_i^{(3,6)}(z) \sin(x+2y) +\frak{f}_i^{(3,7)}(z) \sin(2 x-y)+\frak{f}_i^{(3,8)}(z) \sin(x-2 y),  \quad  \quad \hbox{for} \:\: i=7,8\,,
\end{eqnarray}
For each value of $\gamma$, we now have 64 functions $\frak{f}_i^{(3,\alpha)}(z)$ to solve. As before, each Fourier mode $\alpha$ decouples from all others, so we have 8 independent systems of ODEs to solve (one for each $\alpha$), for each $\gamma$. These ODEs again take the form \eqref{ODEprototype}. The systems corresponding to $\alpha=1,2$, and only these, have $s_i^{(3,\alpha)}\neq 0$, and hence depend on $k^{(2)}$. Therefore, for each of these two $\alpha$'s we solve the equations of motion to find $\frak{f}_i^{(3,\alpha)}(z)$ and $k^{(2)}$. These two systems of equations are independent but $k^{(2)}$ is unique, so we must get the same $k^{(2)}$ in each of them.

There is an important difference between lattices with $\gamma\neq1/2$ and those with $\gamma=1/2$.  The leading contribution to the entropy difference $\Delta \sigma_H$ occurs at order $O(\epsilon^4)$ for the $\gamma\neq1/2$ case, instead of $O(\epsilon^3)$ as in the $\gamma=1/2$ case. We therefore need to continue our computation to   $O(\epsilon^4)$. There are now 13 Fourier modes that are excited. 
Schematically, each of these is of form $\cos({\cal A}^{(4,\alpha)})$, with 
\begin{equation}
{\cal A}^{(4,\alpha)}\in\{0,2x,2y, 4x,4y,x+y,x-y,2(x+y),2(x-y),x+3y,x-3y,3x+y,3x-y\},
\end{equation}
which are precisely those that follow from taking the fourth power of the sum of the two linear-order modes. Solving the associated system of ODEs we find the corresponding Fourier coefficients $\frak{f}_i^{(4,\alpha)}(z)$ for $\alpha=0,\cdots,12$, and $k^{(3)}=0$ (for reasons similar to those that give $k^{(1)}$=0). 

To find the wavevector correction $k^{(4)}$, which is required to compute the thermodynamic quantities up to $O(\epsilon^4)$, we still need to find the metric solutions at order $O(\epsilon^5)$ since $k^{(4)}$ only appears in the equations of motion at this order (in practice it is enough to analyse the $\cos x$ and $\cos y$ Fourier modes since $k^{(4)}$ only appears in the equations associated to this sector). At this order, a total of 18 Fourier modes  are excited. Schematically, each of these is of form $\cos({\cal A}^{(5,\alpha)})$, with 
\begin{equation}
{\cal A}^{(5,\alpha)}\in\{x,y,3x,3y,5x,5y,x+2y,x-2y,2x+y,2x-y,x+4y,x-4y,4x+y,4x-y,2x+3y,2x-3y,3x+2y,3x-2y\},
\end{equation}
which are precisely those that follow from taking the fifth power of the sum of the two linear-order modes. Solving the corresponding ODE system we determine the associated Fourier coefficients $\frak{f}_i^{(5,\alpha)}(z)$ for $\alpha=1,\cdots,18$ and $k^{(4)}\neq 0$.

Having reached the first corrections to the relevant thermodynamic quantities, we do not continue the calculation to higher orders. 

\newpage
\section{Numerical Convergence}
\begin{figure}\centering
\includegraphics[width=.6\textwidth]{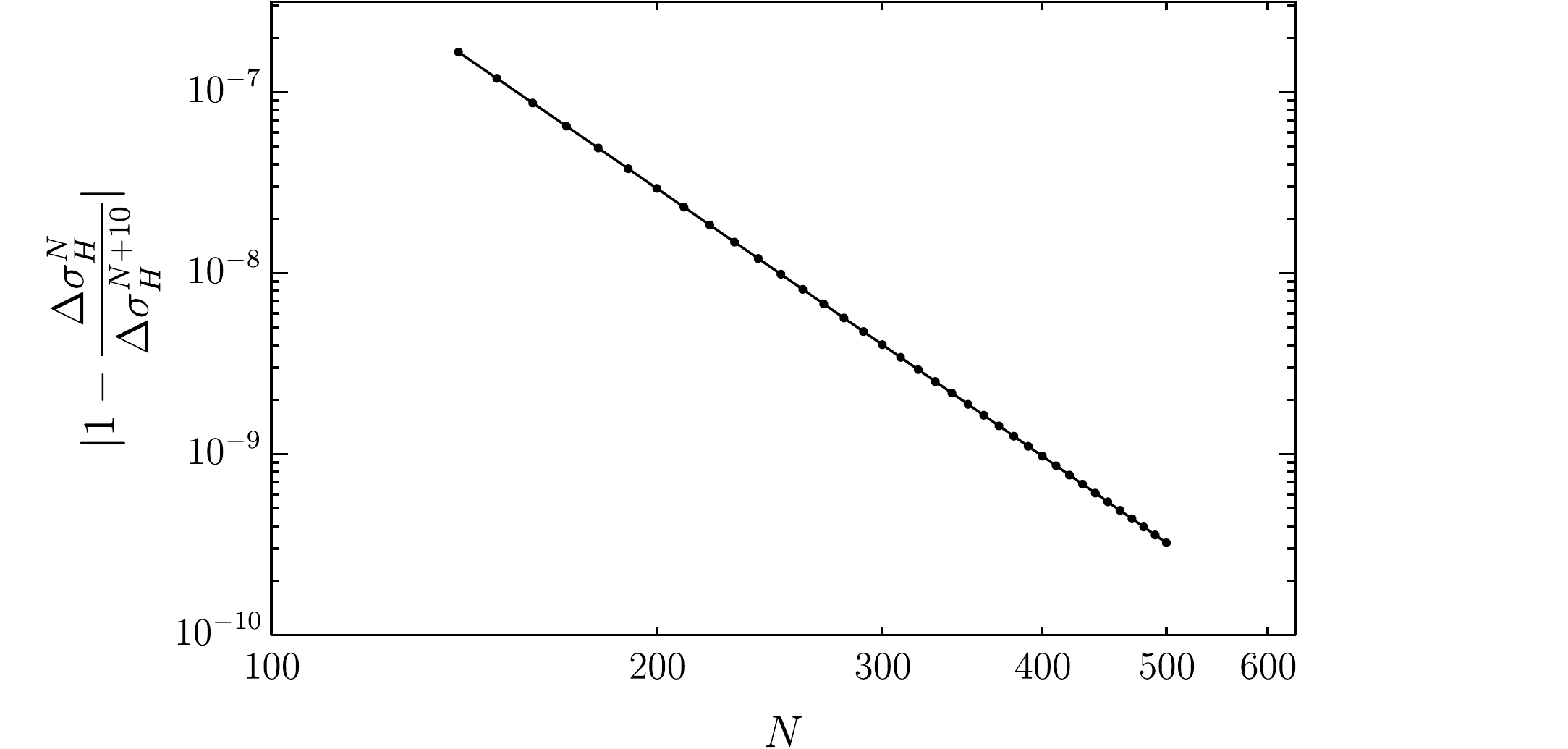}
\caption{Convergence of the $\mathcal O(\epsilon^3)$ coefficient of the entropy difference for $\gamma=1/2$.  The convergence is power-law, which is consistent with the non-smooth decay at infinity.}\label{fig:convergence}
\end{figure}

Here, we show the convergence of our numerical method.  We use a Chebyshev spectral methods which convergences exponentially on smooth functions.  However, the expansion at infinity is not smooth.  We instead find a power-law convergence with grid-size $1/N^{4.91}$, which is consistent with this decay.  A plot of this convergence is shown in Fig. \ref{fig:convergence}.

\bibliography{refs}{}

\begin{thebibliography}{30}
\expandafter\ifx\csname natexlab\endcsname\relax\def\natexlab#1{#1}\fi
\expandafter\ifx\csname bibnamefont\endcsname\relax
  \def\bibnamefont#1{#1}\fi
\expandafter\ifx\csname bibfnamefont\endcsname\relax
  \def\bibfnamefont#1{#1}\fi
\expandafter\ifx\csname citenamefont\endcsname\relax
  \def\citenamefont#1{#1}\fi
\expandafter\ifx\csname url\endcsname\relax
  \def\url#1{\texttt{#1}}\fi
\expandafter\ifx\csname urlprefix\endcsname\relax\def\urlprefix{URL }\fi
\providecommand{\bibinfo}[2]{#2}
\providecommand{\eprint}[2][]{\url{#2}}

\bibitem[{\citenamefont{Gregory and Laflamme}(1993)}]{Gregory:1993vy}
\bibinfo{author}{\bibfnamefont{R.}~\bibnamefont{Gregory}} \bibnamefont{and}
  \bibinfo{author}{\bibfnamefont{R.}~\bibnamefont{Laflamme}},
  \bibinfo{journal}{Phys. Rev. Lett.} \textbf{\bibinfo{volume}{70}},
  \bibinfo{pages}{2837} (\bibinfo{year}{1993}), \eprint{hep-th/9301052}.

\bibitem[{\citenamefont{Dias et~al.}(2009)\citenamefont{Dias, Figueras,
  Monteiro, Santos, and Emparan}}]{Dias:2009iu}
\bibinfo{author}{\bibfnamefont{O.~J.~C.} \bibnamefont{Dias}},
  \bibinfo{author}{\bibfnamefont{P.}~\bibnamefont{Figueras}},
  \bibinfo{author}{\bibfnamefont{R.}~\bibnamefont{Monteiro}},
  \bibinfo{author}{\bibfnamefont{J.~E.} \bibnamefont{Santos}},
  \bibnamefont{and} \bibinfo{author}{\bibfnamefont{R.}~\bibnamefont{Emparan}},
  \bibinfo{journal}{Phys. Rev.} \textbf{\bibinfo{volume}{D80}},
  \bibinfo{pages}{111701} (\bibinfo{year}{2009}), \eprint{0907.2248}.

\bibitem[{\citenamefont{Dias et~al.}(2010)\citenamefont{Dias, Figueras,
  Monteiro, Reall, and Santos}}]{Dias:2010eu}
\bibinfo{author}{\bibfnamefont{O.~J.~C.} \bibnamefont{Dias}},
  \bibinfo{author}{\bibfnamefont{P.}~\bibnamefont{Figueras}},
  \bibinfo{author}{\bibfnamefont{R.}~\bibnamefont{Monteiro}},
  \bibinfo{author}{\bibfnamefont{H.~S.} \bibnamefont{Reall}}, \bibnamefont{and}
  \bibinfo{author}{\bibfnamefont{J.~E.} \bibnamefont{Santos}},
  \bibinfo{journal}{JHEP} \textbf{\bibinfo{volume}{05}}, \bibinfo{pages}{076}
  (\bibinfo{year}{2010}), \eprint{1001.4527}.

\bibitem[{\citenamefont{Hartnett and Santos}(2013)}]{Hartnett:2013fba}
\bibinfo{author}{\bibfnamefont{G.~S.} \bibnamefont{Hartnett}} \bibnamefont{and}
  \bibinfo{author}{\bibfnamefont{J.~E.} \bibnamefont{Santos}},
  \bibinfo{journal}{Phys. Rev.} \textbf{\bibinfo{volume}{D88}},
  \bibinfo{pages}{041505} (\bibinfo{year}{2013}), \eprint{1306.4318}.

\bibitem[{\citenamefont{Dias et~al.}(2014{\natexlab{a}})\citenamefont{Dias,
  Hartnett, and Santos}}]{Dias:2014eua}
\bibinfo{author}{\bibfnamefont{O.~J.~C.} \bibnamefont{Dias}},
  \bibinfo{author}{\bibfnamefont{G.~S.} \bibnamefont{Hartnett}},
  \bibnamefont{and} \bibinfo{author}{\bibfnamefont{J.~E.}
  \bibnamefont{Santos}}, \bibinfo{journal}{Class. Quant. Grav.}
  \textbf{\bibinfo{volume}{31}}, \bibinfo{pages}{245011}
  (\bibinfo{year}{2014}{\natexlab{a}}), \eprint{1402.7047}.

\bibitem[{\citenamefont{Santos and Way}(2015)}]{Santos:2015iua}
\bibinfo{author}{\bibfnamefont{J.~E.} \bibnamefont{Santos}} \bibnamefont{and}
  \bibinfo{author}{\bibfnamefont{B.}~\bibnamefont{Way}},
  \bibinfo{journal}{Phys. Rev. Lett.} \textbf{\bibinfo{volume}{114}},
  \bibinfo{pages}{221101} (\bibinfo{year}{2015}), \eprint{1503.00721}.

\bibitem[{\citenamefont{Emparan and Reall}(2002)}]{Emparan:2001wn}
\bibinfo{author}{\bibfnamefont{R.}~\bibnamefont{Emparan}} \bibnamefont{and}
  \bibinfo{author}{\bibfnamefont{H.~S.} \bibnamefont{Reall}},
  \bibinfo{journal}{Phys. Rev. Lett.} \textbf{\bibinfo{volume}{88}},
  \bibinfo{pages}{101101} (\bibinfo{year}{2002}), \eprint{hep-th/0110260}.

\bibitem[{\citenamefont{Dias et~al.}(2014{\natexlab{b}})\citenamefont{Dias,
  Santos, and Way}}]{Dias:2014cia}
\bibinfo{author}{\bibfnamefont{O.~J.~C.} \bibnamefont{Dias}},
  \bibinfo{author}{\bibfnamefont{J.~E.} \bibnamefont{Santos}},
  \bibnamefont{and} \bibinfo{author}{\bibfnamefont{B.}~\bibnamefont{Way}},
  \bibinfo{journal}{JHEP} \textbf{\bibinfo{volume}{07}}, \bibinfo{pages}{045}
  (\bibinfo{year}{2014}{\natexlab{b}}), \eprint{1402.6345}.

\bibitem[{\citenamefont{Lehner and Pretorius}(2010)}]{Lehner:2010pn}
\bibinfo{author}{\bibfnamefont{L.}~\bibnamefont{Lehner}} \bibnamefont{and}
  \bibinfo{author}{\bibfnamefont{F.}~\bibnamefont{Pretorius}},
  \bibinfo{journal}{Phys. Rev. Lett.} \textbf{\bibinfo{volume}{105}},
  \bibinfo{pages}{101102} (\bibinfo{year}{2010}), \eprint{1006.5960}.

\bibitem[{\citenamefont{Figueras et~al.}(2016)\citenamefont{Figueras, Kunesch,
  and Tunyasuvunakool}}]{Figueras:2015hkb}
\bibinfo{author}{\bibfnamefont{P.}~\bibnamefont{Figueras}},
  \bibinfo{author}{\bibfnamefont{M.}~\bibnamefont{Kunesch}}, \bibnamefont{and}
  \bibinfo{author}{\bibfnamefont{S.}~\bibnamefont{Tunyasuvunakool}},
  \bibinfo{journal}{Phys. Rev. Lett.} \textbf{\bibinfo{volume}{116}},
  \bibinfo{pages}{071102} (\bibinfo{year}{2016}), \eprint{1512.04532}.

\bibitem[{\citenamefont{Figueras et~al.}(2017)\citenamefont{Figueras, Kunesch,
  Lehner, and Tunyasuvunakool}}]{Figueras:2017zwa}
\bibinfo{author}{\bibfnamefont{P.}~\bibnamefont{Figueras}},
  \bibinfo{author}{\bibfnamefont{M.}~\bibnamefont{Kunesch}},
  \bibinfo{author}{\bibfnamefont{L.}~\bibnamefont{Lehner}}, \bibnamefont{and}
  \bibinfo{author}{\bibfnamefont{S.}~\bibnamefont{Tunyasuvunakool}}
  (\bibinfo{year}{2017}), \eprint{1702.01755}.

\bibitem[{\citenamefont{Gubser}(2002)}]{Gubser:2001ac}
\bibinfo{author}{\bibfnamefont{S.~S.} \bibnamefont{Gubser}},
  \bibinfo{journal}{Class. Quant. Grav.} \textbf{\bibinfo{volume}{19}},
  \bibinfo{pages}{4825} (\bibinfo{year}{2002}), \eprint{hep-th/0110193}.

\bibitem[{\citenamefont{Wiseman}(2003)}]{Wiseman:2002zc}
\bibinfo{author}{\bibfnamefont{T.}~\bibnamefont{Wiseman}},
  \bibinfo{journal}{Class. Quant. Grav.} \textbf{\bibinfo{volume}{20}},
  \bibinfo{pages}{1137} (\bibinfo{year}{2003}), \eprint{hep-th/0209051}.

\bibitem[{\citenamefont{Kudoh and Wiseman}(2005)}]{Kudoh:2004hs}
\bibinfo{author}{\bibfnamefont{H.}~\bibnamefont{Kudoh}} \bibnamefont{and}
  \bibinfo{author}{\bibfnamefont{T.}~\bibnamefont{Wiseman}},
  \bibinfo{journal}{Phys. Rev. Lett.} \textbf{\bibinfo{volume}{94}},
  \bibinfo{pages}{161102} (\bibinfo{year}{2005}), \eprint{hep-th/0409111}.

\bibitem[{\citenamefont{Maldacena}(1999)}]{Maldacena:1997re}
\bibinfo{author}{\bibfnamefont{J.~M.} \bibnamefont{Maldacena}},
  \bibinfo{journal}{Int.J.Theor.Phys.} \textbf{\bibinfo{volume}{38}},
  \bibinfo{pages}{1113} (\bibinfo{year}{1999}), \eprint{hep-th/9711200}.

\bibitem[{\citenamefont{Gubser et~al.}(1998)\citenamefont{Gubser, Klebanov, and
  Polyakov}}]{Gubser:1998bc}
\bibinfo{author}{\bibfnamefont{S.~S.} \bibnamefont{Gubser}},
  \bibinfo{author}{\bibfnamefont{I.~R.} \bibnamefont{Klebanov}},
  \bibnamefont{and} \bibinfo{author}{\bibfnamefont{A.~M.}
  \bibnamefont{Polyakov}}, \bibinfo{journal}{Phys. Lett.}
  \textbf{\bibinfo{volume}{B428}}, \bibinfo{pages}{105} (\bibinfo{year}{1998}),
  \eprint{hep-th/9802109}.

\bibitem[{\citenamefont{Witten}(1998)}]{Witten:1998qj}
\bibinfo{author}{\bibfnamefont{E.}~\bibnamefont{Witten}},
  \bibinfo{journal}{Adv. Theor. Math. Phys.} \textbf{\bibinfo{volume}{2}},
  \bibinfo{pages}{253} (\bibinfo{year}{1998}), \eprint{hep-th/9802150}.

\bibitem[{\citenamefont{Aharony et~al.}(2000)\citenamefont{Aharony, Gubser,
  Maldacena, Ooguri, and Oz}}]{Aharony:1999ti}
\bibinfo{author}{\bibfnamefont{O.}~\bibnamefont{Aharony}},
  \bibinfo{author}{\bibfnamefont{S.~S.} \bibnamefont{Gubser}},
  \bibinfo{author}{\bibfnamefont{J.~M.} \bibnamefont{Maldacena}},
  \bibinfo{author}{\bibfnamefont{H.}~\bibnamefont{Ooguri}}, \bibnamefont{and}
  \bibinfo{author}{\bibfnamefont{Y.}~\bibnamefont{Oz}}, \bibinfo{journal}{Phys.
  Rept.} \textbf{\bibinfo{volume}{323}}, \bibinfo{pages}{183}
  (\bibinfo{year}{2000}), \eprint{hep-th/9905111}.

\bibitem[{\citenamefont{Sorkin}(2004)}]{Sorkin:2004qq}
\bibinfo{author}{\bibfnamefont{E.}~\bibnamefont{Sorkin}},
  \bibinfo{journal}{Phys. Rev. Lett.} \textbf{\bibinfo{volume}{93}},
  \bibinfo{pages}{031601} (\bibinfo{year}{2004}), \eprint{hep-th/0402216}.

\bibitem[{\citenamefont{Figueras et~al.}(2012)\citenamefont{Figueras, Murata,
  and Reall}}]{Figueras:2012xj}
\bibinfo{author}{\bibfnamefont{P.}~\bibnamefont{Figueras}},
  \bibinfo{author}{\bibfnamefont{K.}~\bibnamefont{Murata}}, \bibnamefont{and}
  \bibinfo{author}{\bibfnamefont{H.~S.} \bibnamefont{Reall}},
  \bibinfo{journal}{JHEP} \textbf{\bibinfo{volume}{11}}, \bibinfo{pages}{071}
  (\bibinfo{year}{2012}), \eprint{1209.1981}.

\bibitem[{\citenamefont{Rozali and Vincart-Emard}(2016)}]{Rozali:2016yhw}
\bibinfo{author}{\bibfnamefont{M.}~\bibnamefont{Rozali}} \bibnamefont{and}
  \bibinfo{author}{\bibfnamefont{A.}~\bibnamefont{Vincart-Emard}},
  \bibinfo{journal}{JHEP} \textbf{\bibinfo{volume}{08}}, \bibinfo{pages}{166}
  (\bibinfo{year}{2016}), \eprint{1607.01747}.

\bibitem[{\citenamefont{Donos and Gauntlett}(2016)}]{Donos:2015eew}
\bibinfo{author}{\bibfnamefont{A.}~\bibnamefont{Donos}} \bibnamefont{and}
  \bibinfo{author}{\bibfnamefont{J.~P.} \bibnamefont{Gauntlett}},
  \bibinfo{journal}{JHEP} \textbf{\bibinfo{volume}{03}}, \bibinfo{pages}{148}
  (\bibinfo{year}{2016}), \eprint{1512.06861}.

\bibitem[{\citenamefont{Headrick et~al.}(2010)\citenamefont{Headrick, Kitchen,
  and Wiseman}}]{Headrick:2009pv}
\bibinfo{author}{\bibfnamefont{M.}~\bibnamefont{Headrick}},
  \bibinfo{author}{\bibfnamefont{S.}~\bibnamefont{Kitchen}}, \bibnamefont{and}
  \bibinfo{author}{\bibfnamefont{T.}~\bibnamefont{Wiseman}},
  \bibinfo{journal}{Class.Quant.Grav.} \textbf{\bibinfo{volume}{27}},
  \bibinfo{pages}{035002} (\bibinfo{year}{2010}), \eprint{0905.1822}.

\bibitem[{\citenamefont{Figueras et~al.}(2011)\citenamefont{Figueras, Lucietti,
  and Wiseman}}]{Figueras:2011va}
\bibinfo{author}{\bibfnamefont{P.}~\bibnamefont{Figueras}},
  \bibinfo{author}{\bibfnamefont{J.}~\bibnamefont{Lucietti}}, \bibnamefont{and}
  \bibinfo{author}{\bibfnamefont{T.}~\bibnamefont{Wiseman}},
  \bibinfo{journal}{Class.Quant.Grav.} \textbf{\bibinfo{volume}{28}},
  \bibinfo{pages}{215018} (\bibinfo{year}{2011}), \eprint{1104.4489}.

\bibitem[{\citenamefont{Wiseman}(2011)}]{Wiseman:2011by}
\bibinfo{author}{\bibfnamefont{T.}~\bibnamefont{Wiseman}},
  \emph{\bibinfo{title}{{Numerical construction of static and stationary black
  holes}}} (\bibinfo{publisher}{in Black Holes in Higher Dimensions, CUP 2012,
  Ed. Gary T. Horowitz}, \bibinfo{year}{2011}), \eprint{1107.5513},
  \urlprefix\url{http://inspirehep.net/record/920553/files/arXiv:1107.5513.pdf}.

\bibitem[{\citenamefont{Dias et~al.}(2015)\citenamefont{Dias, Santos, and
  Way}}]{Dias:2015nua}
\bibinfo{author}{\bibfnamefont{O.~J.~C.} \bibnamefont{Dias}},
  \bibinfo{author}{\bibfnamefont{J.~E.} \bibnamefont{Santos}},
  \bibnamefont{and} \bibinfo{author}{\bibfnamefont{B.}~\bibnamefont{Way}}
  (\bibinfo{year}{2015}), \eprint{1510.02804}.

\bibitem[{\citenamefont{Harmark and Obers}(2004)}]{Harmark:2004ch}
\bibinfo{author}{\bibfnamefont{T.}~\bibnamefont{Harmark}} \bibnamefont{and}
  \bibinfo{author}{\bibfnamefont{N.~A.} \bibnamefont{Obers}},
  \bibinfo{journal}{JHEP} \textbf{\bibinfo{volume}{05}}, \bibinfo{pages}{043}
  (\bibinfo{year}{2004}), \eprint{hep-th/0403103}.

\bibitem[{\citenamefont{Attems et~al.}(2017)\citenamefont{Attems, Bea,
  Casalderrey-Solana, Mateos, Triana, and Zilhao}}]{Attems:2017ezz}
\bibinfo{author}{\bibfnamefont{M.}~\bibnamefont{Attems}},
  \bibinfo{author}{\bibfnamefont{Y.}~\bibnamefont{Bea}},
  \bibinfo{author}{\bibfnamefont{J.}~\bibnamefont{Casalderrey-Solana}},
  \bibinfo{author}{\bibfnamefont{D.}~\bibnamefont{Mateos}},
  \bibinfo{author}{\bibfnamefont{M.}~\bibnamefont{Triana}}, \bibnamefont{and}
  \bibinfo{author}{\bibfnamefont{M.}~\bibnamefont{Zilhao}},
  \bibinfo{journal}{JHEP} \textbf{\bibinfo{volume}{06}}, \bibinfo{pages}{129}
  (\bibinfo{year}{2017}), \eprint{1703.02948}.

\bibitem[{\citenamefont{Janik et~al.}(2017)\citenamefont{Janik, Jankowski, and
  Soltanpanahi}}]{Janik:2017ykj}
\bibinfo{author}{\bibfnamefont{R.~A.} \bibnamefont{Janik}},
  \bibinfo{author}{\bibfnamefont{J.}~\bibnamefont{Jankowski}},
  \bibnamefont{and}
  \bibinfo{author}{\bibfnamefont{H.}~\bibnamefont{Soltanpanahi}}
  (\bibinfo{year}{2017}), \eprint{1704.05387}.

\bibitem[{\citenamefont{Hanada and Nishioka}(2007)}]{Hanada:2007wn}
\bibinfo{author}{\bibfnamefont{M.}~\bibnamefont{Hanada}} \bibnamefont{and}
  \bibinfo{author}{\bibfnamefont{T.}~\bibnamefont{Nishioka}},
  \bibinfo{journal}{JHEP} \textbf{\bibinfo{volume}{09}}, \bibinfo{pages}{012}
  (\bibinfo{year}{2007}), \eprint{0706.0188}.

\end{thebibliography}
\end{document}